\documentclass[fleqn,usenatbib]{mnras}

\usepackage{newtxtext,newtxmath}
\usepackage{xcolor}

\usepackage[T1]{fontenc}
\usepackage{ae,aecompl}

\usepackage{graphicx}	
\usepackage{amsmath}	
\usepackage{amssymb}	

\title[Absolute colours of TNOs]{Absolute colours and phase coefficients of trans-Neptunian objects: Correlations and Populations}

\author[A. Alvarez-Candal et al.]
{Alvaro Alvarez-Candal,$^{1}$\thanks{E-mail: varobes@gmail.com (AAC)}
Carmen Ayala-Loera,$^{1}$
Ricardo Gil-Hutton,$^{2}$
\newauthor
Jos\'e Luis Ortiz,$^{3}$
Pablo Santos-Sanz,$^{3}$
and
Ren\'e Duffard$^{3}$
\\
$^{1}$Observat\'orio Nacional / MCTIC, Rua General Jos\'e Cristino 77, Rio de Janeiro, RJ, 20921-400, Brazil\\
$^{2}$Departamento de Geof\'isica y Astronom\'ia, Facultad de Ciencias Exactas, F\'isicas y Naturales, Universidad Nacional de San Juan - \\CONICET, Av. Jos\'e I. de la Roza 590 (O), San Juan, J5402DCS, Argentina\\
$^{3}$Instituto de Astrof\'isica de Andaluc\'ia, CSIC, Apt 3004, E-18080 Granada, Spain
}

\date{Accepted XXX. Received YYY; in original form ZZZ}

\pubyear{2019}

\begin{document}
\label{firstpage}
\pagerange{\pageref{firstpage}--\pageref{lastpage}}
\maketitle

\begin{abstract}
The study of the visible colours of the trans-Neptunian objects opened a discussion almost 20 years ago which, in spite of the increase in the amount of available data, seems far from subside. Visible colours impose constraints to the current theories of the early dynamical evolution of the Solar System such as the environment of formation, initial surface composition, and how (if) they were scattered to regions closer to the inner planets.
In this paper we present an updated version of our database of absolute colours and relative phase coefficients for 117 objects. We define the absolute colours as the difference of the absolute magnitudes $H_V-H_R$, and the relative phase coefficient as the difference of the slopes of the phase curves $\Delta\beta$. These were obtained joining our own observations plus data from the literature.
The methodology has been introduced in previous works and here we expand in some interesting results, in particular the strong anti-correlation found between $H_V-H_R$ and $\Delta\beta$, which means that redder objects have steeper phase curves in the R filter, while bluer objects have steeper phase curves in the V filter.
We analyse a series of results published in the literature in view of our database, which is free of phase effects, and show that their statistical meaning is not very strong. We point out that
phase-colouring and observational errors play an important role in the understanding of these proposed relationships.
\end{abstract}

\begin{keywords}
methods: observational -- technique: photometric -- Kuiper belt objects: general
\end{keywords}



\section{Introduction}
The trans-Neptunian population conforms, possibly, the least altered population of minor bodies in the Solar System and the clues of its dynamical and physical-chemical evolution lie among the objects that compose it. Nowadays, nearly 3,000 TNOs (trans-Neptunian objects) are known, including related populations, like the centaurs, their representatives in inner parts of the outer Solar System. In this work we will use the term TNO to refer to all these objects.

Early ideas of the trans-Neptunian belt regarded it as dynamically cold and with similar physical properties \cite[e.g.][]{ferna1980}. Nevertheless, with the increase of the available observational data, it became clear that the belt showed a large degree of heterogeneity: visible colours ranging from nearly solar to extremely red \cite[e.g.][and references therein]{LuuJe1996,dores2008,peixi2015,tegle2016}, and surface composition that varies from displaying absorption bands (methane, molecular nitrogen, or water ices, and a few other traces) to being almost featureless within the signal-to-noise ratio \citep{baruc2011,brown2012}. The large dispersion of properties is interpreted in terms of the TNOs dynamical evolution since the Solar System formation \cite[e.g.][]{nesvo2018}, the chemical evolution of the surfaces \cite[e.g.][]{strazz1991,hudso2008}, and loss of volatile off the surfaces \citep{schal2007}.

Most of the available data on TNOs comes from photometric studies, which is the fastest and easiest observational way to characterise a large population of minor bodies through large data-sets. Early studies pointed towards a possible bi-modal distribution of their visible colors \citep{tegle1998}, although not all works coincided with this view \cite[for instance][]{LuuJe1996}. The former work pointed towards an issue with the observational errors reported. The so-called ``colours controversy'' is still on-going, while \cite{peixi2015} point to an apparent bi-modal distribution only for Centaurs and TNOs fainter than $H_V\approx7$, \cite{tegle2016} find it as a property of the whole population. Moreover, the apparent bi-modality in the visible colours, especially B-R, is posited as the responsible for the colour distribution of other small bodies populations, such as the Trojans \citep{wongb2016}. This is still highly debatable \citep[e.g.,][]{jewit2018}.

Many of the publications used colours obtained at one single phase angle, $\alpha$\footnote{The angular distance between the Earth and the Sun as seen from the object.}, relying on the fact that, as $\alpha<2$ deg for TNOs, these magnitudes were close enough to zero phase to be corrected by phase effects, or, if they were corrected, many used average values of the phase coefficient, $\beta$ (see equation~\ref{eq:1} below).

Since 2012 we have been collecting V and R magnitudes, ours and from the literature, in order to create an ensemble of absolute magnitudes, i.e., free of any possible phase-related effect, in both filters as described in \citet[][paper1]{alcan2016} and \citet[][paper2]{ayala2018}. 
In paper2 we defined the absolute colour as $H_V-H_R$ and the relative phase coefficient $\Delta\beta=\beta_V-\beta_R$. The former is the colour of the object free of any phase effect, while the second represents the phase-colouring tendency of the object, i.e., it represents the phase coefficient of the colour phase curve, with opposite sign (equation~\ref{eq:1}). The results from paper1 and paper2 can be summarised as:
\begin{itemize}
    \item TNOs show a large dispersion of phase coefficients ($\beta_{\lambda}$). Therefore, care must be taken when assuming an average value.
    \item {The behaviour of the phase curves apparently does not depend on the albedo of the objects.}
    \item $H_V-H_R$ strongly correlate with $\Delta\beta$. The correlation is independent of surface composition (including albedo), size, and location in the outer Solar System.
\end{itemize}
In this paper we continue the work started in paper2, aiming at exploring the possible cause of the strong correlation between $H_V-H_R$ and $\Delta\beta$. We also look into many of the relations proposed for the colour distribution of TNOs in the literature. 

This manuscript is organised as follows: in Sect. \ref{database} we summarily describe the updates to our database since paper2, including new observations of 6 objects. Section \ref{processing} describes the processing of the data, which has been slightly improved with respect to paper1 and paper2, while in Sect. \ref{results} we present the results that are finally discussed in Sect. \ref{discussion}. We draw our conclusion in the last section.

\section{The database}\label{database}
In this section we describe updates to our database since paper2. First we present new observations of 6 objects, while afterwards we describe new data collected from the literature.

\subsection{Observations}\label{obs}
The new observations were performed at the SOAR 4-m telescope in Cerro Pach\'on (Chile)\footnote{http://www.ctio.noao.edu/soar/} using Brazilian allocated time. Observations were carried out with the SOI camera\footnote{http://www.ctio.noao.edu/soar/content/soar-optical-imager-soi} using the V and R filters and a $2\times2$ binning, providing a scale plate of $\sim0.15$ arcsec per pixel. Data reduction was performed using daily calibration files (flat-field and bias) following standard methods with IRAF. Standard stars from \citet{lando1992} and \citet{clela2013} were used to compute the zero points. Average extinction coefficients and colour terms were used to calibrate the instrumental magnitudes. A detailed description of the reduction process can be found in paper1. Table \ref{table:obs} shows the V and R magnitudes of the observed objects, as well as their heliocentric ($r$) and topocentric ($\Delta$) distances, and phase angle at the time of observations. The ephemeris were obtained from JPL's horizons\footnote{https://ssd.jpl.nasa.gov/horizons.cgi}.
\begin{table*}
	\centering
	\caption{New observations. The table lists the observed objects, the night of observations, the V and R magnitudes, and the heliocentric and topocentric distances, as well as the phase angle.} 
	\label{table:obs}
	\begin{tabular}{lcccccc} 
		\hline
		Object & Date & V & R & $r$ (AU)& $\Delta$ (AU)& $\alpha$ (deg)\\
		\hline
   2000OK67    & 2014-08-04&  21.96 (0.18)& 21.30 (0.17)& 40.1447&  39.3539 &   0.9215\\  
  2001QY297    & 2014-08-04&  21.78 (0.33)& 21.22 (0.20)& 43.5343&  42.5432 &   0.2950\\  
   2007JK43    & 2014-08-04&  20.43 (0.05)& 19.66 (0.04)& 23.6786&  23.2651 &   2.2586\\  
   2007RW10    & 2014-08-04&  20.95 (0.09)& 20.16 (0.07)& 28.7173&  28.3226 &   1.8805\\  
  2010JJ124    & 2014-08-29&  21.05 (0.57)& 21.17 (0.11)& 23.6983&  23.4975 &   2.4014\\  
Teharonhiawako & 2014-08-04&  21.64 (0.13)& 21.67 (0.28)& 45.1292&  44.2257 &   0.5974\\  
		\hline
	\end{tabular}
\end{table*}

\subsection{Data collected from the literature}
Since the final editorial stages of paper2 we became aware of new articles, or older works that had escaped our attention, that included V and/or R magnitudes. In this work we include data from the references listed in Table \ref{table:new_data}.
\begin{table*}
	\centering
	\caption{New data collected from literature.}
	\label{table:new_data}
	\begin{tabular}{ll} 
		\hline
		Reference & Objects \\
		\hline
		
\cite{LuuJe1996} & 1993FW 1993RO 1993SC 1994ES2 1994EV3 1994JS 1994TB 1995DA2 1995DB2 1995DC2 1995QY9  \\
                 & 1995WY2 2002GX32 Albion Chiron Hylonome Nessus Pholus \\

\cite{laclu2006} & 1996TO66 1996TS66 1998SN165 Chaos    \\  

\cite{bagnu2006} & 2002VE95      \\   

\cite{ortiz2007} & Makemake \\   

\cite{belsk2008} & Eris   \\    

\cite{tegle2016} & 1996GQ21 1998VG44 2000GN171 2001KB77 2001QF298 2002AW197 2002KX14 2002KY14 2002MS4  \\
                 & 2002PQ15 2002QX47 2002TC302 2002UX25 2002VE95 2002VR128 2002WC19 2002XV93 2003AZ84  \\
                 & 2003FY128 2003UR292 2003UY117 2003UZ117 2003VS2 2003WL7 2004EW95 2004GV9 2005RO43   \\
                 & 2005UJ438 2006SX368 2006XQ51 2007RG283 2007RH283 2007TJ422 2007TK422 2007UM126 2007VH305    \\
                 & 2008FC76 2008QD4 2008SJ236 2008UZ6 2008YB3 2009YF7 2009YG19 2010BK118 2010NV1 2010RM64   \\
                 & 2010TH 2010VZ98 2011ON45 2012UT68 2012VU852013UL10  2013XZ8 2014ON6 Amycus Echeclus Eris \\
                 & Makemake Orcus Rhiphonos Sedna Varda  \\  
		\hline
	\end{tabular}
\end{table*}\\

The new data includes 139 data points for 92 different objects. Therefore, our updated database includes over 2,400 entries, each representing an individual night, comprising 340 objects. {The increase in the number of data points per object produces a denser coverage of the phase curves which in turns increases the precision of our results (see Appendix \ref{appendixb}).}

\section{Data processing}\label{processing}
From our database we selected all objects that have at least three observations in, at least, three different phase angles, in both filters, not necessarily simultaneous, as we intend to compute $H_V-H_R$. In total we have 122 selected objects. We follow the same procedure described in our two earlier papers (paper1 and paper2), therefore, we will not repeat it here step by step. Instead we summarise it in the next two paragraphs, giving some more detail to minor improvements applied for objects with no light-curve information.

\subsection{Objects with known light-curves}\label{sideltam}
We use the rotational information, only peak-to-peak amplitudes ($\Delta m$), from \cite{thiro2012} and \cite{benec2013}. Whenever there were two different values reported for the same object we choose the largest one. If only an upper limit was reported we assumed it as $\Delta m$. In total we have a list of 135 objects with an estimation of $\Delta m$, among them 78 are in our database.

For these 78 objects we computed $H_V$, $\beta_V$, $H_R$, and $\beta_R$ following equation~(\ref{eq:1})
\begin{equation}\label{eq:1}
M_{\lambda}(1,1,\alpha) = H_{\lambda}+\beta_{\lambda}\times\alpha,
\end{equation}
where $M_{\lambda}(1,1,\alpha) = M_{\lambda}-5\log{(r\Delta)}$ is the reduced magnitude, $M_{\lambda}$ is the observed apparent magnitude, $H_{\lambda}$ is the absolute magnitude, $\beta_{\lambda}$ is the phase coefficient, and $\alpha$ is the phase angle. We weighted each $M_{\lambda}(1,1,\alpha)$ by the corresponding error $\sigma_{M_{\lambda}(1,1,\alpha)}=\sigma_{M_{\lambda}}$.

\begin{figure}
	\includegraphics[width=6cm,angle=90]{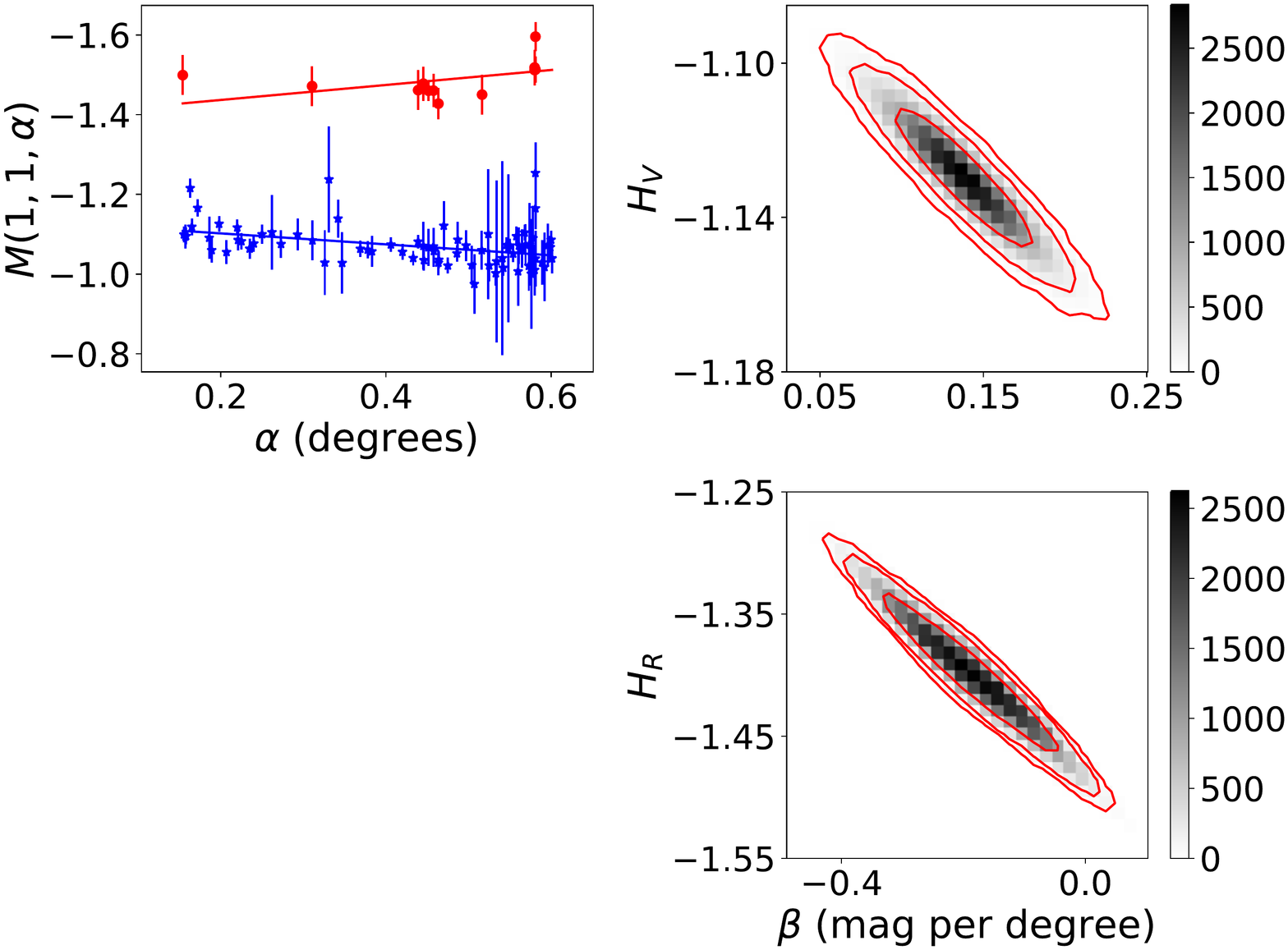}
    \caption{Left panel: Phase curves for the TNO Eris. In blue data in V filter, while in red data in R filter. The continuous lines represent the average values of the 100,000 solutions. Right panel: The phase space covered by the solutions. The contours indicate where the 68\%, 95\%, and 99.7\% of the solutions are contained.}
    \label{fig:example}
\end{figure}

As explained in paper1 and paper2, we generated 100,000 different solutions of equation~(\ref{eq:1}) by making the substitution $M_{\lambda}(1,1,\alpha)\rightarrow M_{\lambda}(1,1,\alpha)+rand_i\times\Delta m/2$, where $rand_i$ is drawn randomly from an uniform distribution $\in~[-1,1]$. We use a flat distribution due to its simplicity and because it avoids selecting preferential rotational phases, for instance a $\sin(x)$ distribution will favour multiples of $\pi/2$ rotational phases and, due to its shape, it can produce bi-modal solutions in the phase space $\beta_{\lambda}$ vs. $H_{\lambda}$. Note that, differently from our previous publications, we use the semi-amplitude instead of the full amplitude as it overestimated the errors by allowing too many solutions with points below (above) the minimum (maximum), without substantially changing the average values, which we assigned as the nominal values (see Appendix \ref{appendixa}). One example of the processing can be seen in Fig. \ref{fig:example} for (136199) Eris.

\subsection{Objects with unknown light-curves}\label{nodeltam}
The remaining 39 objects do not have an estimated $\Delta m$. To provide a good estimation of their absolute magnitude and phase coefficients we modified the previous version of our algorithm that plainly assumed the median value of the distribution for all of them. In this version we make a two-step estimation. First, we used the median value of the distribution ($\Delta m= 0.15$) and proceeded as described above, but using only 100 realisations of equation~(\ref{eq:1}) to produce a first guess of the solutions, which we call $H_{\lambda}'$ and $\beta_{\lambda}'$.

\begin{figure}
	\includegraphics[width=6cm,angle=90]{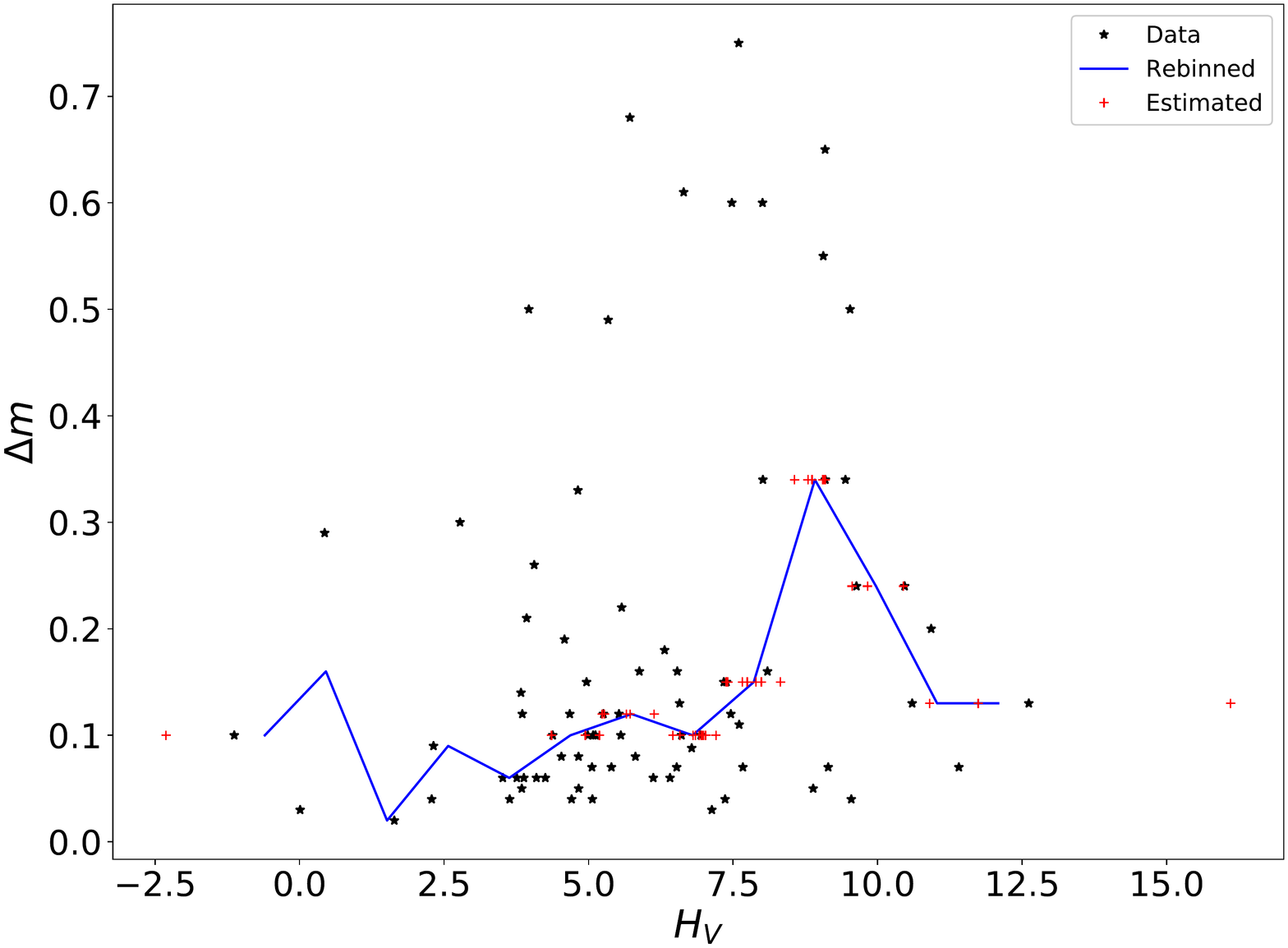}
    \caption{Absolute magnitude in the V band vs. light-curve amplitude. The black asterisks indicate values taken from the literature (or their upper limits). The blue line shows the median-rebbined data, and the red plus symbols the assigned values of $\Delta m$ for objects with no reported light-curves.}
    \label{fig:deltam_hv}
\end{figure}

Using the $H_V$ computed in Sect \ref{sideltam} we created a median curve in the space $\Delta m$ vs. $H_V$ (blue curve in Fig. \ref{fig:deltam_hv}) by binning the space into 13 equally-sized bins between the maximum and minimum value of $H_V$. Now, using $H_V'$ for the 39 objects with no known rotational amplitude, we estimated in which bin they should appear and, accordingly, assigned the median value of $\Delta m$ in that bin. The idea behind this approach is to include in our $\Delta m$ estimation any possible size-dependent effect. {For instance, \cite{duffard2009} explored the possibility that objects with different values of $\Delta m$ have their light-curves dominated by different mechanisms: Large objects tend to have $\Delta m<0.15$ and light-curves possibly dominated by albedo markings, while the smaller TNOs have $\Delta m>0.15$, and their light-curves might be dominated by shape.} We used the median value within each bin, rather than the average, because it is less sensible to outliers. {It is interesting to note that the median curve does not follows any clear pattern, although more data are necessary to risk any interpretation.}
We then re-run the code with the full 100,000 solutions and proceeded as in the Sect. \ref{sideltam}.

Finally, we discarded all objects with values $|\beta_{\lambda}|>1.5$ mag per degree (5). We obtained $H_V-H_R$ and $\Delta\beta$ for 117 objects, 13 more than in paper2\footnote{Note that due to a typo in paper2 we mentioned 106 objects with colours, when in fact there are only 104.}. All figures can be found at {\tt http://extranet.on.br/alvarez/paper3/}.

\section{Results}\label{results}
In this work we focus on the analysis of the relationships between the absolute colours, the relative phase coefficients, and the orbital elements\footnote{ftp://ftp.lowell.edu/pub/elgb/astorb.html}, geometric albedos\footnote{http://public-tnosarecool.lesia.obspm.fr/} and dynamical classes\footnote{http://www.johnstonsarchive.net/astro/tnoslist.html} of the objects in our sample. 
A sample of the measured absolute colours can be seen in Table \ref{table:colors_db}, the rest of the data are available online at {\tt http://extranet.on.br/alvarez/paper3/}. Table \ref{table:colors_db} reports object in the first column, $H_V-H_R$ and $\Delta\beta$, and respective errors, in columns 2 and 3. The number of data points used are reported in columns 4 and 5 (V and R, respectively). Column 6 shows the value of $\Delta m$ used, if Flag=0 in column 7, then we used a value reported in the literature, otherwise $\Delta m$ was estimated as explained in Sect. \ref{nodeltam}. {Finally, columns 8 and 9 show the minimum and maximum phase angles of observation in the V and R filters, respectively.}
\begin{table*}
	\centering
	\caption{Sample of absolute colours and relative phase coefficients and respective errors in parenthesis. The full version can be found on-line.}
	\label{table:colors_db}
	\begin{tabular}{lcccccccc} 
		\hline
		Objects & $H_V-H_R$ & $\Delta\beta$ (mag per deg) &$N_V$ & $N_R$ & $\Delta m$ & Flag & $\alpha_{min}$ (V/R, deg)& $\alpha_{max}$ (V/R, deg) \\
		\hline 
     1993FW&      0.368 	(0.124)& 	  0.269     (0.111)&   4 &  5&  0.10 &1&  0.0017/0.0017&  1.0895/1.0895\\
     1993RO&      0.399 	(0.110)& 	  0.197     (0.172)&   5 &  6&  0.34 &1&  0.0462/0.0462&  0.9575/0.9575\\
     1993SB&      0.294 	(0.148)& 	  0.066     (0.118)&   5 &  4&  0.15 &1&  0.245 /0.8558&  1.5296/1.5296\\
     1993SC&      0.661 	(0.015)& 	  0.045     (0.012)&   9 &  7&  0.04 &0&  0.1497/0.1497&  1.4672/1.4672\\
    1994EV3&      0.559 	(0.107)& 	 -0.052     (0.158)&   4 &  5&  0.15 &1&  0.222 /0.222 &  0.8661/0.8661\\
    1994JQ1&      0.280 	(0.066)& 	  0.751     (0.089)&   5 &  5&  0.10 &1&  0.0285/0.0285&  0.9513/0.9513\\
     1994TB&      0.791 	(0.222)& 	 -0.073     (0.144)&  10 &  8&  0.34 &0&  0.2855/0.8191&  1.7416/1.7416\\
		\hline
	\end{tabular}
\end{table*}

To test the correlations we used the Spearman rank-order correlation test \citep{Spear1904} adapted to take into account the errors in the data, as in paper1. In a nutshell, we perform 100,000 different test by selecting each data point from a normal distribution with central value equal to our nominal and standard deviation equal to the error. The space of solutions are displayed as 2-dimensional histograms where it can be seen whether the data has large excursions, implying that errors could change the suspected correlation (and therefore that we cannot reject the null hypothesis), while if the solutions are concentrated the test holds against the errors. As a remainder, we consider a correlation as plausible if $|r_s|>0.5$ and $P_{r_s}\approx0$.

\subsection{Orbital elements and dynamical classes}
For simplicity, and to avoid using groups of small number of objects, we separated our sample into three groups: (i) {\bf resonants}: gathering all objects trapped into mean motion resonances (32 objects, of which 25 are Plutinos); (ii) {\bf classical}: including all Cubewanos, 32 objects, regardless whether they are dynamically ``hot'' or ``cold'', plus the members of the Haumea group (7 objects); and finally {\bf cds}: including all the rest, centaurs (32 objects), scattering objects (8 objects), and others (6 objects).

We will not attempt to search for every possible correlation between the different parameters, instead we focus only on interesting relations reported by diverse authors and analyse them in light of our results.

\subsubsection{Colour and perihelion distance}
\begin{figure}
	\includegraphics[width=6cm,angle=90]{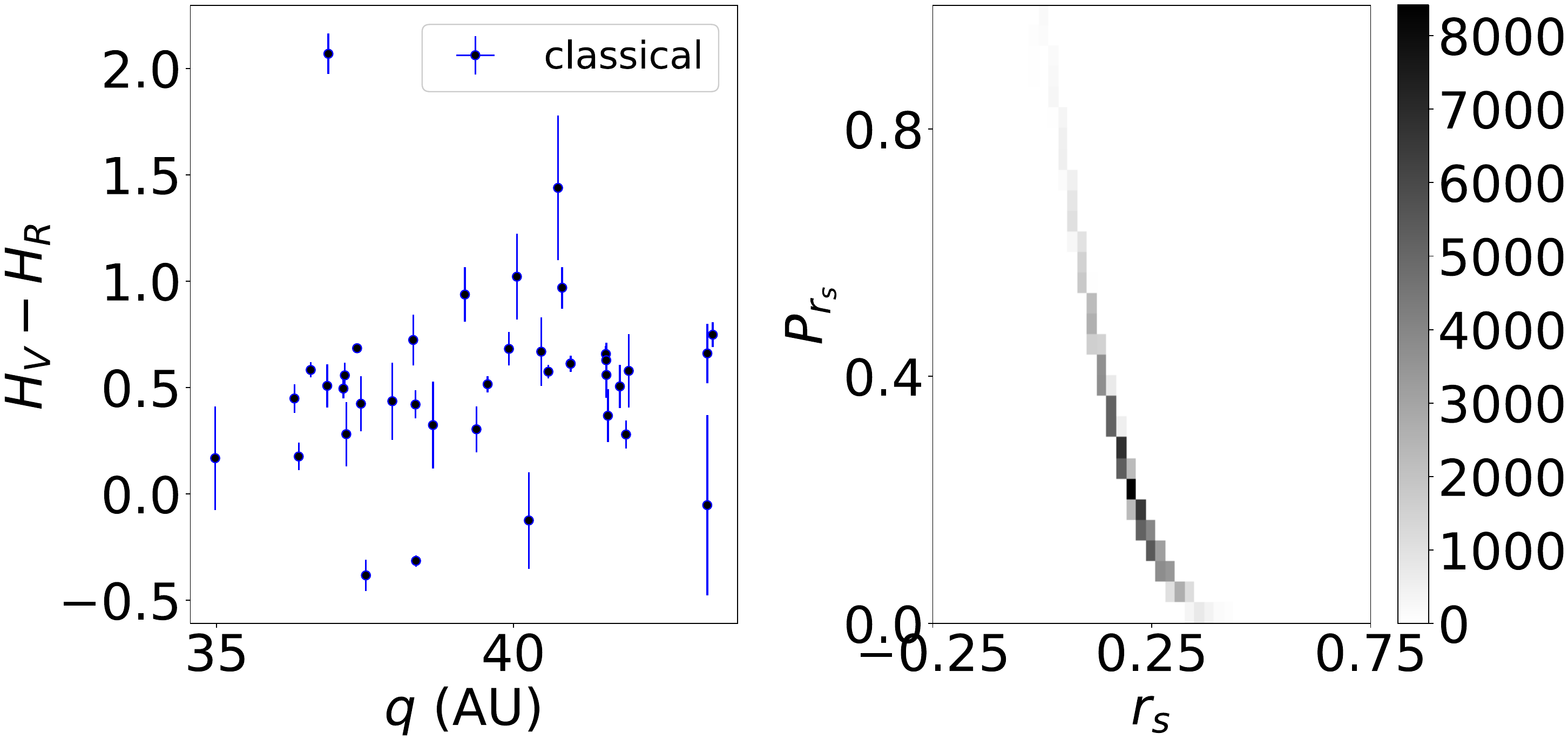}
    \caption{Left: Perihelion distance vs. $H_V-H_R$ of classical objects. Right: Phase space covered by the Spearman test of correlation when considering the errors in the data. The large excursions show that the data do no correlate. Nominal values: $r_s = 0.21$, $P_{r_s}=0.192$.}
    \label{fig:per_cla}
\end{figure}

In the early 2000's it was proposed the existence of a clump of red TNOs with low eccentricity orbits and perihelion distance larger than 40 AU \citep{tegle2000}. We checked with our data and found no clear evidence of a red clump among the whole population. Interestingly, when analysing the classical group, we detect that the colour seems to increase, on average, until about 41 AU, when it appears a clump of objects with $H_V-H_R\approx0.4$ and $q\approx 42$ AU (Fig. \ref{fig:per_cla}). Note that this is not influenced by the Haumea group that have perihelion distances below 39 AU. Nevertheless, we stress that this ``increase'' lacks of any statistical significance and it is only driven by visual inspection, as shown by the right panel in Fig. \ref{fig:per_cla}.

\subsubsection{Colour and inclination}
\begin{figure}
	\includegraphics[width=6cm,angle=90]{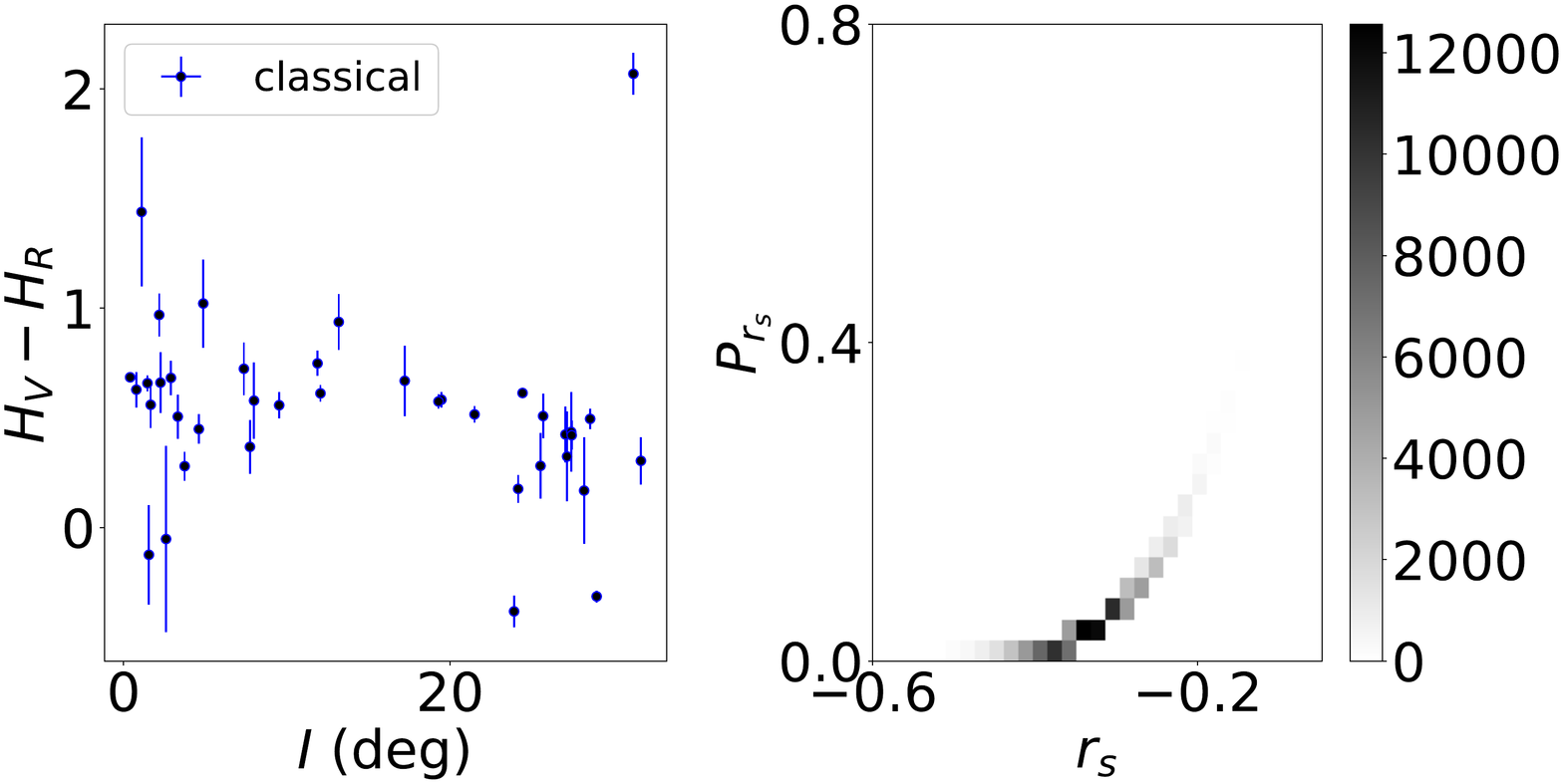}
    \caption{
    Left: Inclination vs. $H_V-H_R$ of classical objects. Right: Phase space covered by the Spearman test of correlation when considering the errors in the data. The large excursions show that the data is weakly correlated. Nominal values:
    $r_s = -0.38$, $P_{r_s}=0.017$.}
    \label{fig:inc_cla}
\end{figure}

It was recognised by \cite{dores2008} that the TNOs seem to hold a correlation between visible colours and inclinations  with a significance over $5\sigma$, probably due to an identified correlation among the classical TNOs. We searched for this correlation and found only weak evidence of correlation for the classical population (Fig. \ref{fig:inc_cla}).

Interestingly, from the figure it is apparent that at low inclinations ($<10$ deg) there exists a large dispersion of colours, ranging from red to blue, while at larger inclinations there seems to be only bluer objects. This is at odds with the usual view of a red clump of low-inclination objects and a larger distribution of colours in the high inclination population \citep[e.g.][]{peixi2008}. Noteworthy, this observational difference has been widely used as one of the observational constrains to separate the hot and cold populations in the classical trans-Neptunian belt and relate it with the dynamical evolution of the objects \citep{gomes2003}.

{Recently, \cite{marset2018} pointed out that the excited TNO population (all TNOs minus classical with $I\leq5$ deg, the Haumea group, objects in retrograde orbits, and objects with Tisserand parameter with respect to Jupiter less than 3) show a clear bi-modal distribution of colours (their Fig. 1). We used the same constrains and built a similar plot (64 objects, Fig. \ref{fig:inc_exc}), where, in spite of a few objects with extreme values, it is no possible to conclude that there are two different behaviours. There seems to be slight difference in the average values of $H_V-H_R$ for objects above and below $I=15$ deg. Nevertheless, the difference is not statistically significant: the mean $H_V-H_{R_{(I\leq15~{\rm deg)}}}=0.61\pm0.43$, while the mean $H_V-H_{R_{(I>15~{\rm deg)}}}=0.58\pm0.56$. Also, the eye might be misled by a hint of an increase in colour up to 15 deg, and then a decrease for larger inclinations. But these are not statistically significant relations.
\begin{figure}
	\includegraphics[width=6cm,angle=90]{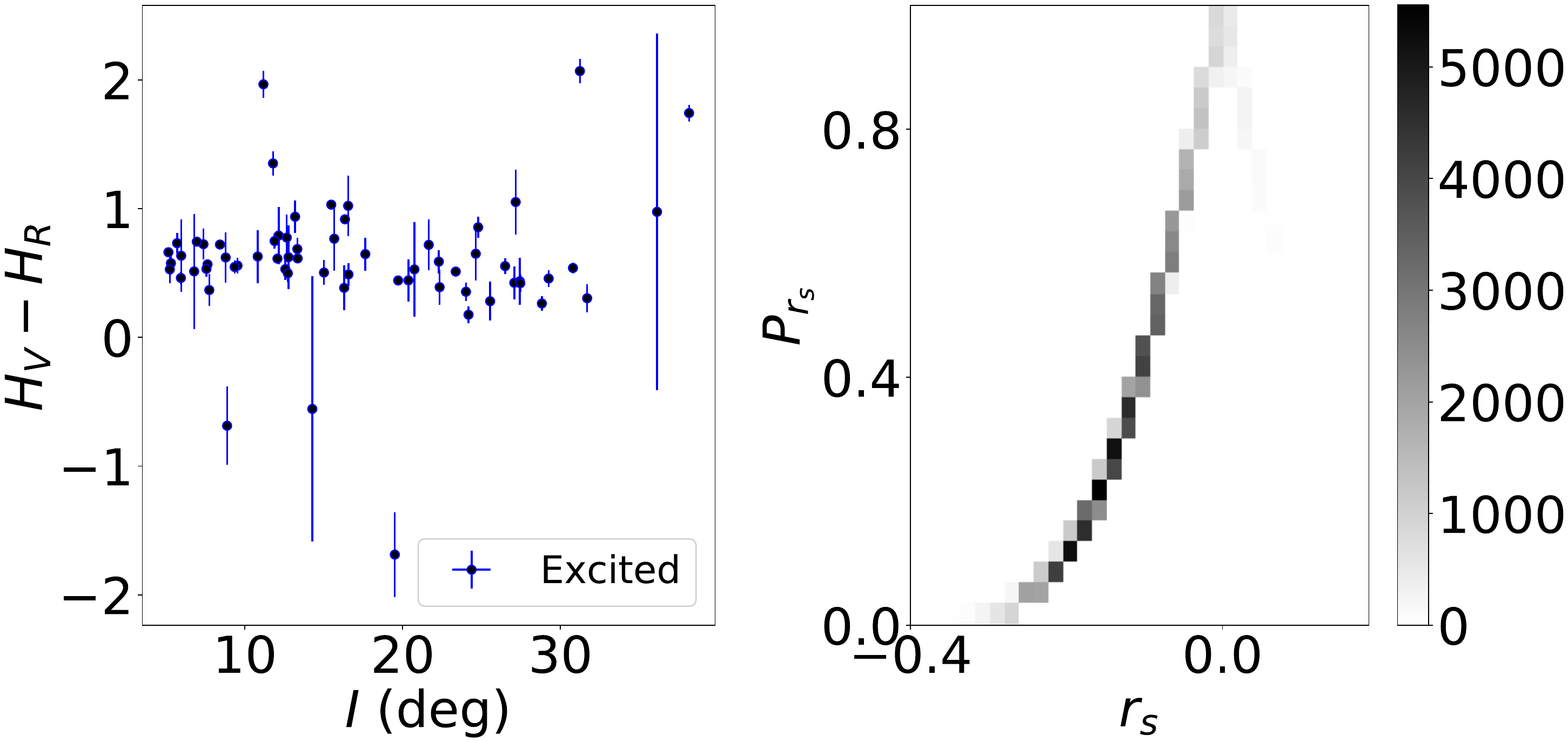}
    \caption{
    Left: Inclination vs. $H_V-H_R$ of the excited TNOs. Right: Phase space covered by the Spearman test of correlation when considering the errors in the data. The large excursions show that the data is not correlated. Nominal values:
    $r_s = -0.14$, $P_{r_s}=0.253$.}
    \label{fig:inc_exc}
\end{figure}}

\subsection{Colour and $\Delta\beta$}
\begin{figure}
	\includegraphics[width=6cm,angle=90]{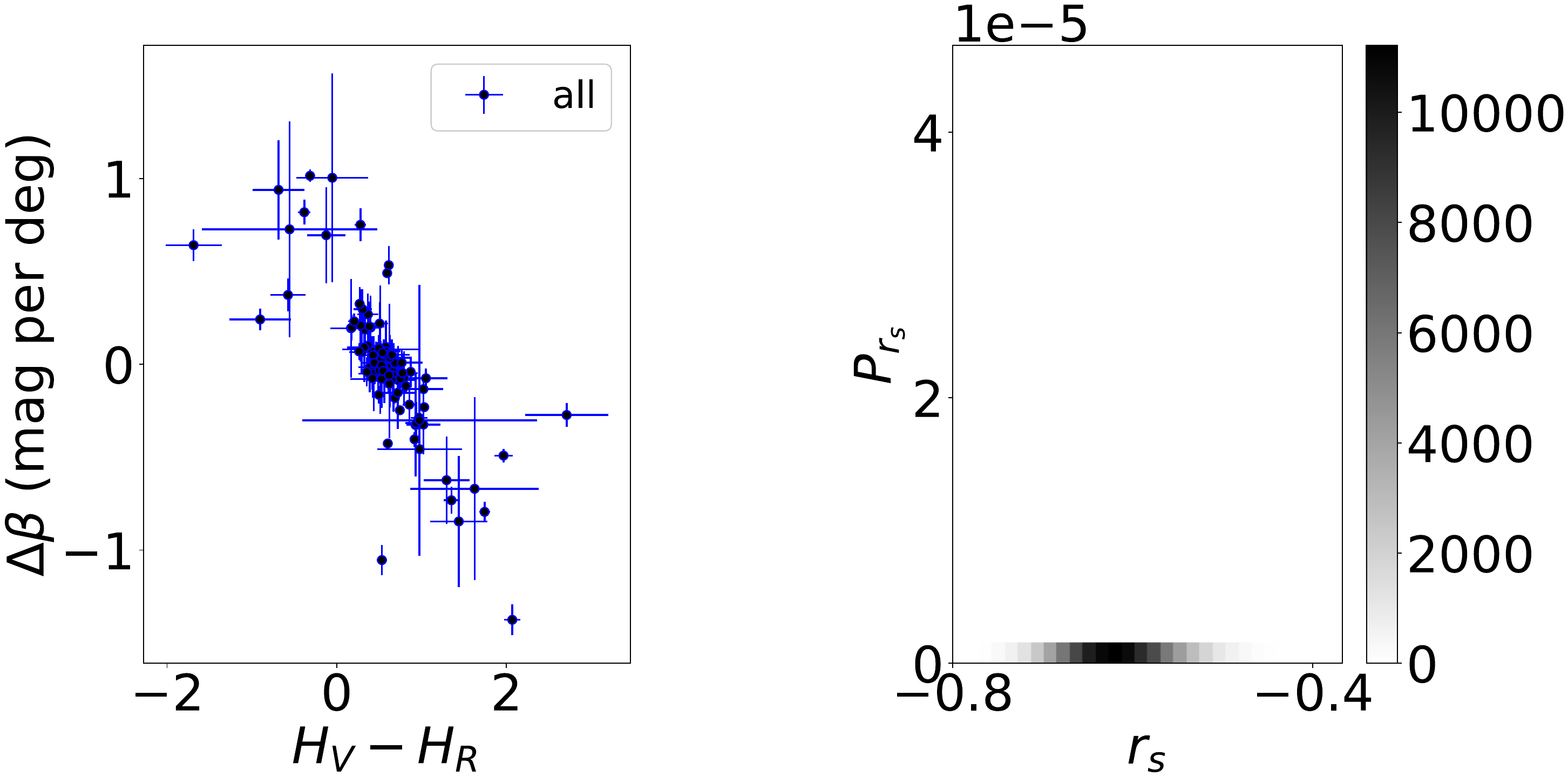}
    \caption{
    Left: $H_V-H_R$ vs. $\Delta\beta$ for all our sample. Right: Phase space covered by the Spearman test of correlation when considering the errors in the data. The very small space covered by the 100,000 solution shows that the correlation is strong. Nominal values:
    $r_s = -0.79$, $P_{r_s}=10^{-25}$.}
    \label{fig:color_beta}
\end{figure}

In paper2 we found that TNOs follow a very strong anti-correlation between $H_V-H_R$ and $\Delta\beta$. Here we update these results including the new data described above (see Fig. \ref{fig:color_beta}).
As described in the paper2, the correlation holds when considering different size ranges and bins in semi-major axis. We searched for correlations within each of the three dynamical groups defined above finding also very strong anti-correlations (see Table \ref{table:col_bet}).

{We performed two tests to check if the quality of our results had any impact on the relation: (i) We only used objects with more than 4 points in V and R, (ii) we only used objects with $\sigma_{H_{\lambda}}<0.05$. In both cases the relation holds, for all populations and the combined data-set, although with a larger $P_{r_s}$.
Also, as pointed out in paper1 and paper2, some phase curves do not follow the expected increase in brightness towards smaller $\alpha$. In fact, several objects have at least one $\beta_{\lambda}$ negative. We checked that the correlation holds even if we use only the ``well-behaved'' objects (73 with positive values of both $\beta_{\lambda}$).}

\begin{table}
	\centering
	\caption{$H_V-H_R$ vs. $\Delta\beta$ Spearman correlations for all dynamical samples considered in this work.}
	\label{table:col_bet}
	\begin{tabular}{lcc} 
		\hline
		Population & $r_s$ & $P_{r_s}$ \\
		\hline
All            & -0.78     &  $10^{-25}$\\  
Resonant       & -0.76     &  $4\times10^{-7 }$\\  
Classical      & -0.83     &  $8\times10^{-11}$\\  
cds            & -0.76     &  $7\times10^{-10}$\\
		\hline
	\end{tabular}
\end{table}
These results clearly point towards a property of the whole population, independently of location in the Solar System, surface temperature, composition (and/or albedo), and size. In paper2 we theorised that the anti-correlation could be due to microscopical properties of the surfaces, but we did not go any further.

\subsubsection{Photometric modelling}

{Aiming at exploring possible explanations for the anti-correlation it is possible to compare these results with a theoretical model. The obvious option is the official IAU magnitude system, which is based on the three-parameter model developed by \cite{muino2010}, but it is a fit to the observations which in turn also depends on tabulated functions and in this case it is necessary to compare with a model that allow some kind of relation between physical and model parameters. Then, the older model proposed in \cite{lumme1981}  would be better adapted to these requirements and is of particular interest to us due to its equation 62b that relates the multiple-scattering factor to the geometric albedo, and therefore, it is possible to tune it to different wavelengths.}

We will use the relation:
\begin{equation}\label{eq:2}
D[km]\propto10^{-H/5}\times p^{-0.5}
\end{equation}
to relate the ratio of albedos in V and R to the absolute colour via:
\begin{equation}\label{eq:3}
10^{(H_V-H_R)/5}\propto(p_V/p_R)^{0.5}
\end{equation}
and compute numerical phase curves to be compared with our observational results.

\begin{figure}
	\includegraphics[width=6cm,angle=90]{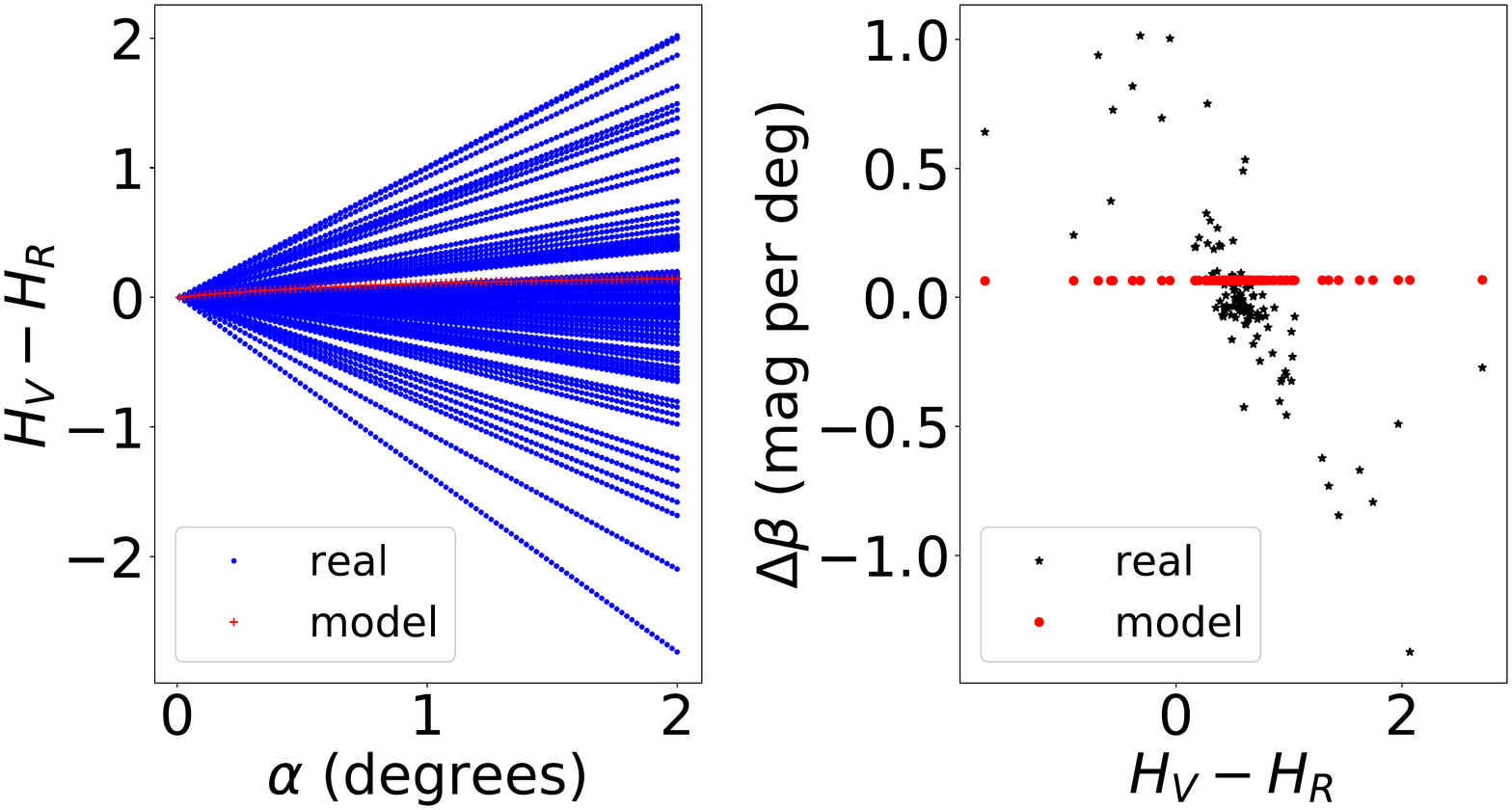}
    \caption{Comparisons between the observational results and the results obtained from the photometric model. Left: In blue are shown the colour phase curves of our objects (normalised to $H_V-H_R=0$ for clarity) and in red the same curves if they followed the models. Right: In black asterisks are marked our data, without errorbars for clarity, in red appear the space covered by the model for the same colours.}
    \label{fig:model}
\end{figure}

Figures \ref{fig:model} show the results of the modelling. As it stands out, the phase space covered by the modelled data does not reproduce the observational results.
The right panel in Fig. \ref{fig:model} reproduces Fig. \ref{fig:color_beta}, without error bars, and overplots the results of running the \citet{lumme1981} model obtaining phase curves in filters V and R with the constrain of the measured values of $H_V$ and $H_R$. To compare the values of $\Delta\beta$ from the modelled phase curves with ours we made a linear fit to the models. In red points are shown the result where it is clear that the phase space is not even close to be recovered. In the left panel of the figure we compare the variation of colour with phase angle between the observed data (blue points) and the modelled data (red points). The modelled data behaves as suspected, the colours redden with increasing phase angle (phase reddening), while the real data, on the other hand, do not necessarily redden with increasing phase angle. Moreover, most of the objects become bluer with increasing phase angle, the opposite of the expected behaviour.

\subsection{Colour distributions}
\begin{figure}
	\includegraphics[width=6cm,angle=90]{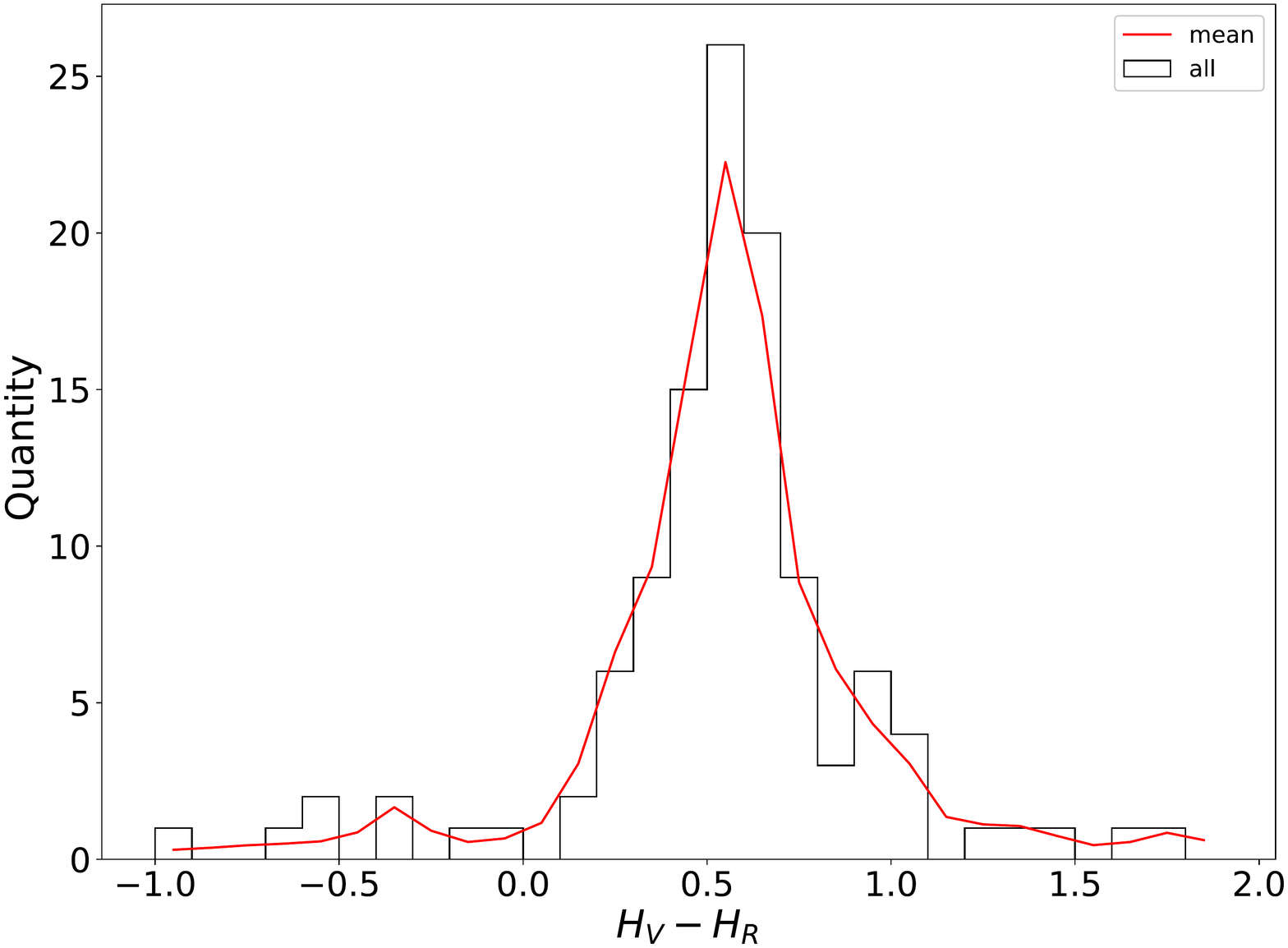}
    \caption{Histogram showing $H_V-H_R$ of the complete sample (black line). In red are shown the resulting histogram when errorbars are taken into account (see text).}
    \label{fig:histo_all}
\end{figure}

We created histograms of the colour distribution of our sample, not with the intention to check every possible histogram and distribution, but to look into reported results. The histograms were made in bins of 0.1 mag in width. For the complete data-set the histogram is shown in black line in Fig. \ref{fig:histo_all}. In the histogram, it could be possible do discern two modes, a strong peak at $H_V-H_R\approx0.5$ and a smaller one at about $H_V-H_R\approx0.9$. This could be though as evidence for the proposed bi-modality of colours in the literature \citep[e.g.][]{tegle2016}. Nevertheless, we feel that the histogram does not reflect the effect of the errors in the colour distribution.

To take into consideration the errors in the data, we proceeded similarly as mentioned elsewhere in the text: We created 100,000 different distributions by drawing the colours from a normal distribution with a mean value equal to the nominal $H_V-H_R$ and with a standard deviation equal to $\sigma_{H_V-H_R}$. We then computed the average over each bin and this is shown as a red line in Fig. \ref{fig:histo_all}. This procedure smooths out the histogram and, for instance, it makes disappear any secondary mode. Nevertheless, caution should taken whenever there are few objects (1 or 2) because this procedure could create features that are not real (for instance the peak at $H_V-H_R\approx-0.25$ in the red line distribution).

\begin{figure}
	\includegraphics[width=6cm,angle=90]{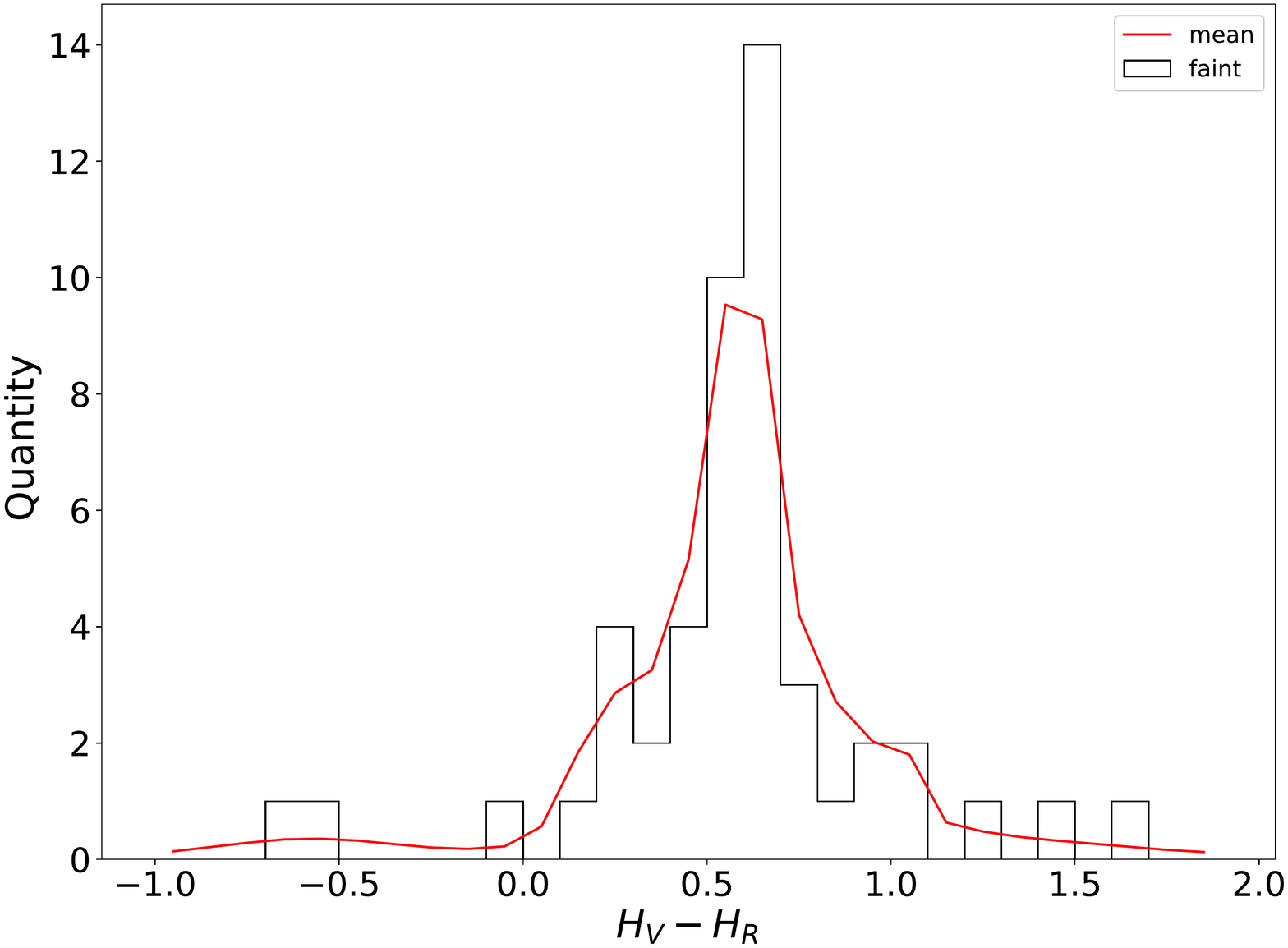}
    \caption{Histogram showing $H_V-H_R$ of the objects fainter than $H_V = 7$ (49 objects, black line). In red are shown the resulting histogram when errorbars are taken into account (see text).}
    \label{fig:histo_faint}
\end{figure}

Perhaps one of the most interesting distribution to look into is that of objects fainter than $H_V\approx7$, which could show a bi-modal distribution of colours \citep[Fig. 13 in][]{peixi2015}. In Fig. \ref{fig:histo_faint} we show all objects fainter than $H_V = 7$, processed in the same way as the previous plot. 
It is not clear the existence of a bi-modal distribution, neither in the nominal histogram nor in the ``averaged'' one. A very clear mode is present at $H_V-H_R\approx0.6$ and a very wide distribution of colours becomes apparent in the averaged histogram.

\section{Discussion}\label{discussion}
We split the discussion into three parts to tackle different issues raised by our results, some of them were already mentioned in paper1 and/or paper2, but here we discuss them more in detail.\\

\noindent(1)
In \cite{ayala2018} we identified a strong relation between the absolute colour and the relative phase coefficients for a sample of 106 objects, this relationship holds regardless of location in the outer Solar System and size. In this work we confirm that it is present among the different dynamical classes, therefore, it must be a property of the particles forming the surface of the bodies, rather than composition, perhaps due to the size-particle distribution and/or inter-particle spacing.

We used the model presented in \cite{lumme1981} aiming at interpreting, at last partially, our results. The idea behind this attempt was to check if, by obtaining phase curves in different wavelengths, the multiple scattering term could be providing an effect of shadowing different in V and R filters. 
Unfortunately, and although in principle the model does allow to use different albedos, which we approximate via the absolute colour (see equation~\ref{eq:3}), the phase space covered by the modelled phase curves does not permit solutions as extreme as we obtained from the observational data. Even considering that some outliers in our results might be interfering, for instance the extreme blue and red colours in Fig. \ref{fig:color_beta}, the bulk of the sample still shows the strong relation and it is not even closely sampled by the modelled results.

{The phase angle range covered for TNOs is mostly within the region of the opposition effect, which has two main sources: shadow-hiding \citep{hapke1963} and coherent back-scattering \citep{muino1989}. Both effects should contribute to an increase in the magnitude close to $\alpha = 0$ deg, although with different angular widths, wavelength dependencies, and strengths \citep[Sect. 8.H. in][]{hapke1993}.

Our data show that redder objects have steeper phase curves in R filter than in V filter. If we assume that the slope of the phase curves is somehow associated to the angular width of the opposition effect, OE, then steeper~slopes $\Rightarrow$ thinner~widths, and the OE is wider in the V filter than in the R filter for the redder objects. This effect is the opposite than expected if multiple-scattering, and therefore coherent back-scattering, were the dominant source of OE \citep{hapke1993}. Figure \ref{fig:model} already showed that using the multiple-scattering factor does not cover the phase space by our results. Furthermore, \cite{bagnulo2008} showed that TNOs have two different polarimetric behaviours and that these seem at odds with the coherent back-scattering scenario. Therefore, one possibility is that single-scattering is dominant in the surfaces of TNOs. 
The OE could also be related to the parity of the electric field within the scatterers having an effect similar to the OE produced by multiple-scattering \citep{muinonen2007}. Perhaps, multiple-wavelength phase curves is the way to disentangle both effects. {Could the albedo have some effect? As shown in our previous works (see Table 1 in paper1 and Table 3 in paper2) our data do not support the existence of a relationship between albedo and the behaviour of the phase curves. Nevertheless, it is important to bear in mind that it is possible that the uncertainties in our data, and in the albedos, are hiding any such relation. Perhaps with the increase in the number of stellar occultations by TNOs it would be possible to decrease the uncertainties in the albedos unveiling some relation. }

If single-scattering is the dominant mechanism, \cite{hapke1963} showed that less compact surfaces have sharper phase curves, therefore, redder objects show different compactions at different wavelengths: more compact in V than in R, while for more neutral objects, the compaction is similar. Therefore, the $H_V-H_R$ vs. $\Delta\beta$ relation points towards spacing between the scatterers on the surface, which could be large, as suggested for Eris \citep{belsk2008}.

An interesting step forward would be to combine our data with polarization phase curves for all our objects. \cite{gilhutton2017} showed that there exists a relation between the distance, $d$, of the scatterers and the phase angle where the minimum of the polarization curve happens ($\alpha_{min}$) and it is possible to infer the value of $d$. Nevertheless, their results were obtained for main belt asteroids and might not be directly applicable to TNOs because $\alpha_{min}>5$ deg, values that are only reached by a Centaurs in our sample. 

Another possibility is that equation~(\ref{eq:3}) is wrong due to a different diameter of the object in V and R which can be a result of some kind of activity or caused by a collapsed and re-sublimated atmosphere similar to that suggested by \cite{sicar2011} for Eris. }\\

\begin{figure}
	\includegraphics[width=\columnwidth]{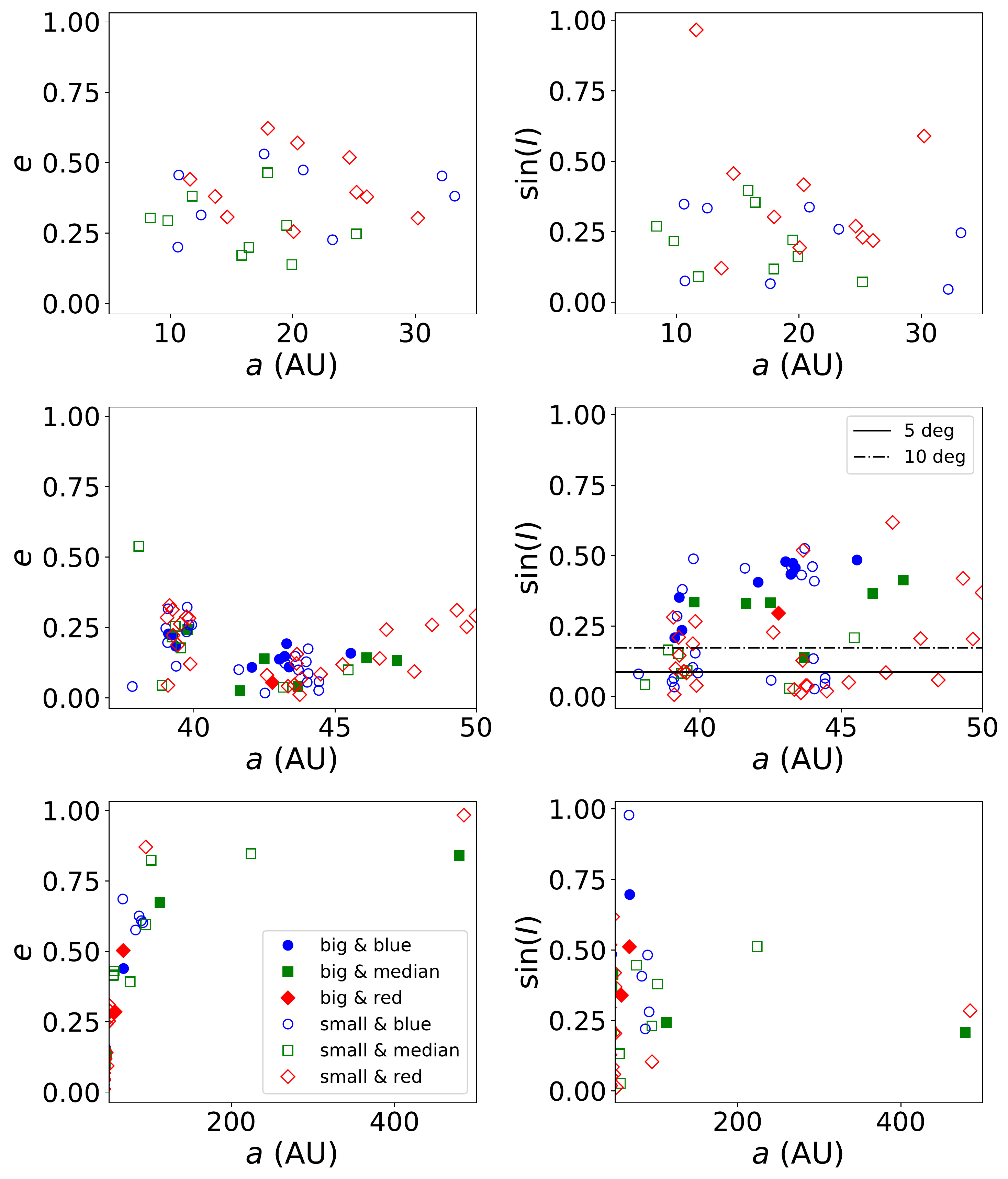}
    \caption{Orbital distribution of the observed objects. Open symbols indicate objects fainter than $H_V=4.5$ (small), while filled symbols indicate objects brighter than $H_V=4.5$. Blue circles indicate objects $H_V-H_R<0.9\times$ the median colour, green squares indicate colours within 10\% of the median, while red diamonds indicate objects redder than $1.1\times$ the median colour.}
    \label{fig:elements}
\end{figure}

\noindent(2)
In Fig. \ref{fig:elements} we show the distribution of our objects in the space of orbital elements. We have chosen to represent them graphically using the median value of $H_V-H_R=0.567$, as seen in the distribution shown in Fig. \ref{fig:histo_all}. Objects bluer than 10\% from the median are shown in blue, within 10\% of the median in green, while redder than 10\% of the median are shown in red. Filled and open symbols separate objects brighter and fainter than $H_V=4.5$, respectively ($D\sim500$ km, see paper2). The top panels show the Centaur space, the middle panels the main trans-Neptunian belt, and the bottom panels the extended trans-Neptunian region.
As seen in the figure, there is no obvious concentration of objects, in any colour interval, in any emplacement. With the only exception of an apparent overabundance of red objects at low inclination, but it should be noted that there are also blue objects at low inclinations, as seen in the middle right panel of Fig. \ref{fig:elements}.

\begin{figure}
	\includegraphics[width=\columnwidth]{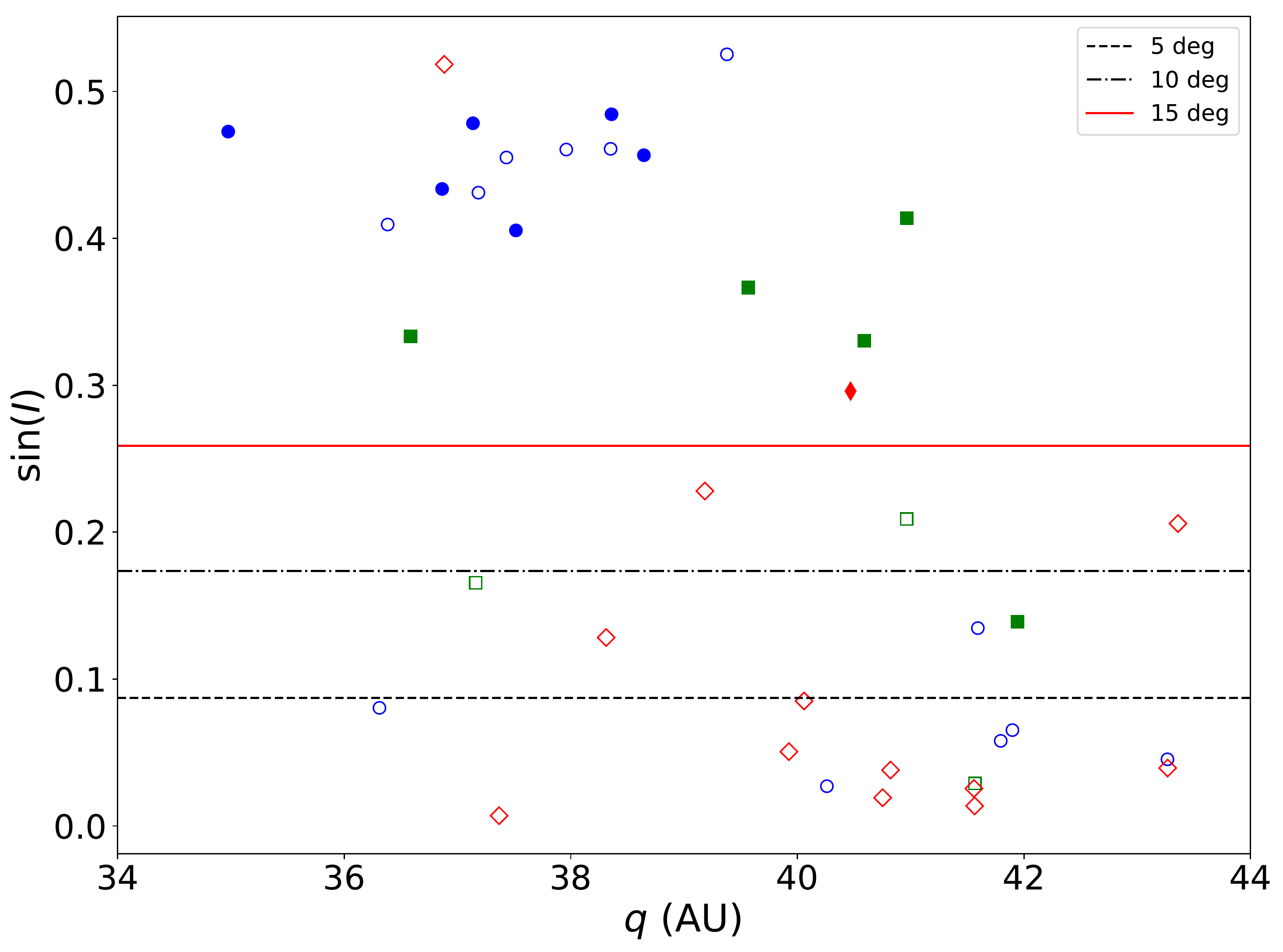}
    \caption{Perihelion distance ($q$) vs. $\sin{I}$ for our defined classical group. The colour scheme and symbol filling follow Fig. \ref{fig:elements}. The horizontal lines show $I=5$ deg (black dashed), $I=10$ deg (black dot-dashed), and $I=15$ deg (red continuous).}
    \label{fig:qicla}
\end{figure}

\begin{figure}
	\includegraphics[width=6cm,angle=90]{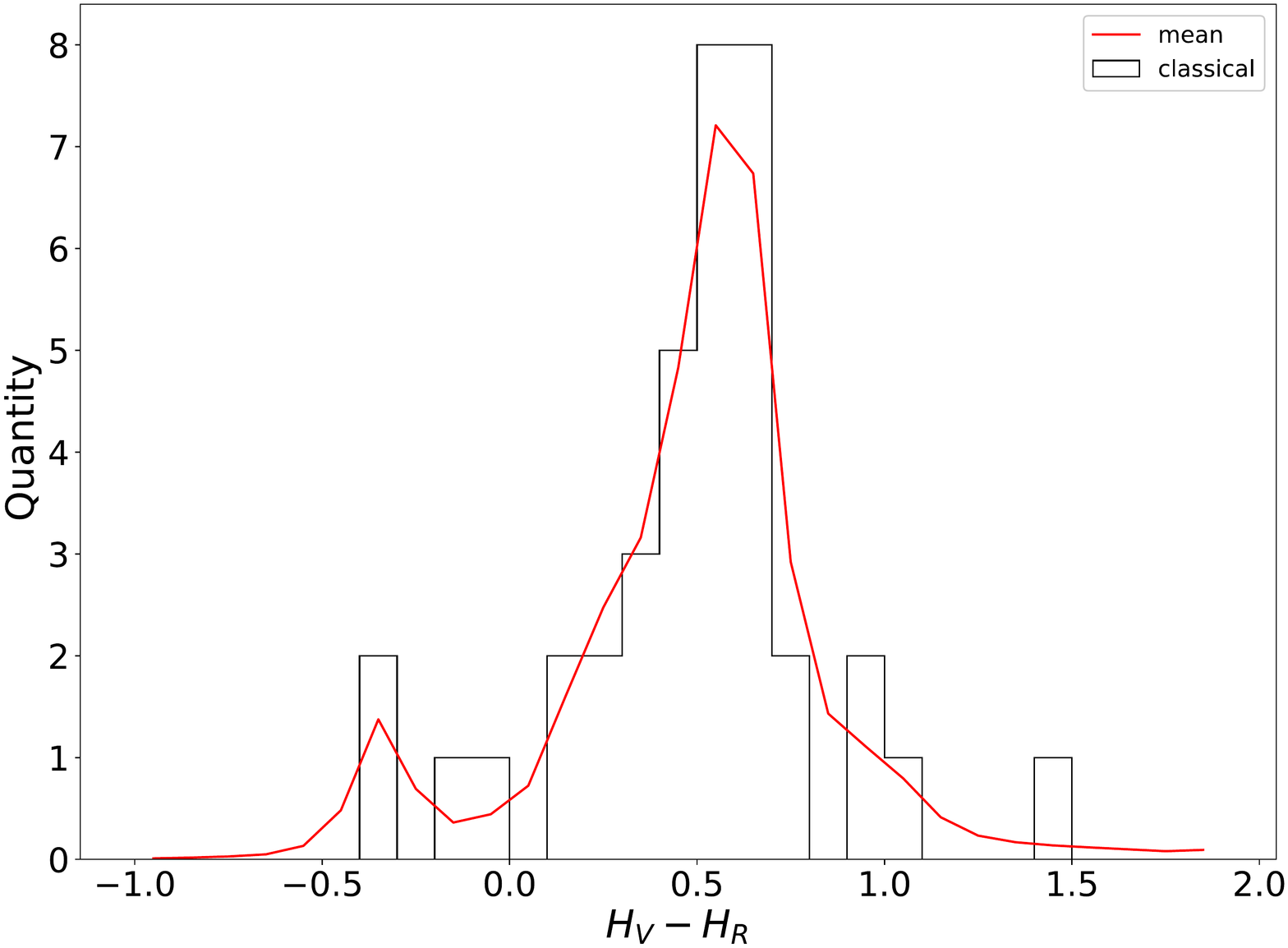}
    \caption{Histogram showing $H_V-H_R$ of the classical group (black line). In red are shown the resulting histogram when errorbars are taken into account (see text).}
    \label{fig:histo_cla}
\end{figure}

Figures \ref{fig:per_cla} and \ref{fig:inc_cla} show that there are a few interesting structures in the TNO region, although not as strong as previously reported. In Fig. \ref{fig:qicla} we show data for only the classical group, following the same symbol and colour definitions as in Fig. \ref{fig:elements}. There, we show that there is a mix of objects with different colours spread over the whole space. But, there seems to be a distinction at about $I=15$ deg, marked with a red continuous line in the figure. Below 15 deg, there are 21 objects, 11 of which (52 \%) are redder than $1.1\times$ the median $H_V-H_R$, while at $I>15$ deg there are 18 objects of which 12 (67 \%) are bluer than $0.9\times$ the median $H_V-H_R$. Therefore, even if we see a large degree of mixture in colours, we can still identify a larger cold classical population of redder and fainter objects at low inclinations. Noteworthy, the histogram of colours of the classical group does not show any hint of bi-modal distribution, but it does show a wide mode (Fig. \ref{fig:histo_cla}).

Perhaps only a curiosity, but classical objects with high-$q$ and high-$I$ are lacking in our dataset.
None of the other two populations, resonant nor csd, show any significant structure.\\

\noindent(3)
We explored the possible bi-modal distribution of colours reported in diverse publications. For instance, recent results \citep{wongb2017} posit that the possible bi-modal colour distribution of the TNOs is due to the chemical evolution of H$_{2}$S in different locations of Solar System prior to the great instability that reshaped the architecture of the early Solar System \citep{nesvor2012}. The chemical evolution of H$_{2}$S produced a very red residue due to irradiation on the surfaces, producing the red and the very red populations \citep[as defined by][]{wongb2017}. This model was put forward aimed at explaining the bi-modal colour distribution reported for Trojan asteroids \citep{wongb2016}. But the model does not, for instance, consider that the colours of Neptune's trojans cannot be reproduced \citep{jewit2018} and these could have the same origin as Jupiter's \citep[e.g.,][]{lykaw2010}. Moreover, \cite{wong2019} recently found out that the simple H$_2$S model does not explain their UVB observations of Trojans. It is also worth noting that recently it was discovered the first ultra-red Neptune trojan \citep{lin2019}, but this is still far from being a confirmation of an hypothetical bi-modality. We created diverse histograms and found no clear evidence of bi-modality on any of them. Therefore, our results point towards a more or less continuous distribution of colours.

Most works that report correlations and/or bi-modal distributions of colours, usually report them in wavelength intervals wider than covered by the $V-R$ colours we use. For instance, \cite{tegle1998,peixi2015,tegle2016} use also B filter data, \cite{wongb2017} uses $g-i$ colours, or \cite{perna2010} uses the taxonomies of TNOs defined using several different colours. On the other hand, \cite{peixi2003}, though reporting BVR colours, showed that the bi-modal evidence was stronger in their $V-R$ data. Furthermore, most spectra of TNOs are quite linear in the visible spectral range \citep[e.g.,][]{alcan2008,fornasier2009} and it should be possible to, at least, have the hints of the bi-modal behaviour using any colour within that range. The main difference between these works and ours is that our colours are not affected by phase as they are not snapshots taken at one epoch, or averaged over close-in-time nights, and, as shown in this work, and also in paper1 and paper2, using observations at a single $\alpha$ could be misleading.

Furthermore, it would be interesting to compare our results with the absolute colours of other related population, such as Trojans (both Jupiter's and Neptune's), Hilda asteroids, and D-type asteroids in general. And certainly increase, not only our database of V and R magnitudes, but also add other filters to start comparing photo-spectra at $\alpha=0$ deg.

\section{Conclusions}\label{conclusion}
Our database is composed of a large, and quite heterogeneous, set of reported magnitudes in the V and R filters, from different telescopes, instruments, and observing conditions. The set of filters are not all the same, with slightly different central wavelengths and band widths. Nevertheless, in most of the cases, for any given object, the data are consistent, therefore we trust our results to be valid. Furthermore, in paper1 we compared our $H_V$ with values from other groups finding good agreement, thus, supporting our methodology. Our errors tended to be larger than those other works, but this was (at least partially) due to an overestimation of the light-curve amplitude effect on the phase curves (see the Appendix \ref{appendixa}).

Our main results can be summarised as:
\begin{itemize}
    \item We obtained $H_V-H_R$ and $\Delta\beta$ for 117 objects, an increase of about 10\% with respect to our previous paper.
    \item The strong anti-correlation between these quantities appears not only in the complete database, but also when considering different dynamical classes, and it is very likely related to micro properties of the surfaces.
    \item It is no longer possible to assume that the colour of an object observed at any random phase angle is representative of the colour at opposition, as large changes can happen.
    \item Our colours do not support a strong bi-modal-colour thesis, instead they point towards a continuum. 
\end{itemize}

\section*{Acknowledgements}

The authors appreciate the comments and suggestions sent by an anonymous referee that helped improve this manuscript.

\noindent
{\bf Facilities:}
Partially based on observations obtained at the Southern Astrophysical Research (SOAR) telescope, which is a joint project of the Minist\'erio da Ci\^encia, Tecnologia, Inova\c c\~ao e Comunica\c c\~oes (MCTIC) da Rep\'ublica Federativa do Brasil, the U.S. National Optical Astronomy Observatory (NOAO), the University of North Carolina at Chapel Hill (UNC), and Michigan State University (MSU).

\noindent
{\bf Funding:} 
AAC acknowledges support from FAPERJ (grant E26/203.186/2016) and CNPq (grants 304971/2016-2 and 401669/2016-5).
CAL thanks CNPq's support (studentship 141784/2015-6).
RGH gratefully acknowledge support by CONICET through PIP 112-201501-00525, and San Juan National University by a CICITCA grant for the period 2018-2019.
JLO thanks support from grant AYA2017-89637-R.
JLO, PSS, and RD acknowledge financial support from the State Agency for Research of the Spanish MCIU through the ``Center of Excellence Severo Ochoa'' award for the Instituto de Astrof\'isica de Andaluc\'ia (SEV-2017-0709).
PSS acknowledges financial support by the European Union's Horizon 2020 Research and Innovation Programme, under Grant Agreement no 687378, as part of the project ``Small Bodies Near and Far'' (SBNAF).

\noindent
{\bf Software:}
IRAF is distributed by the National Optical Astronomy Observatory, which is operated by the Association of Universities for Research in Astronomy (AURA) under a cooperative agreement with the National Science Foundation.
https://www.python.org/.
https://www.scipy.org/.
https://www.libreoffice.org/.
Matplotlib \citep{hunte2007}.




\bibliographystyle{mnras}
\bibliography{biblio} 




\appendix
{\section{Precision vs. number of point in the phase curves}\label{appendixb}
The increase in the number of points per object should produce an increase in the precision of the values of $H$ and $\beta$. We performed a simple test aimed at confirming this statement.

We assumed a generic value of $\beta=0.1$ mag per deg and $H=10.0$. We used this values in equation~(\ref{eq:1}) and generated 20 random phase angles within [0,2] deg to obtain 20 values of $M(1,1,\alpha)$. We added the effect of a possible light-curve, assuming $\Delta m=0.25$, by generating random $\delta m$ from an uniform distribution in $[-\Delta m/2,\Delta m/2]$. Each $M(1,1,\alpha)$ has an error extracted from the positive wing of a normal distribution with $\sigma =0.2$.

We processed the phase curves in the same way as explained in Sect. \ref{sideltam}, first using only 3 of the points, then 4 points, and so on.  Notice that we made sure to select randomly the points in the phase curves. Figure \ref{fig:presicion} shows the results: the errors in $H$ and $\beta$ from our simulation clearly decrease with increasing number of points used to create the phase curves, especially when having more then five data points.
\begin{figure}
	\includegraphics[width=6cm,angle=90]{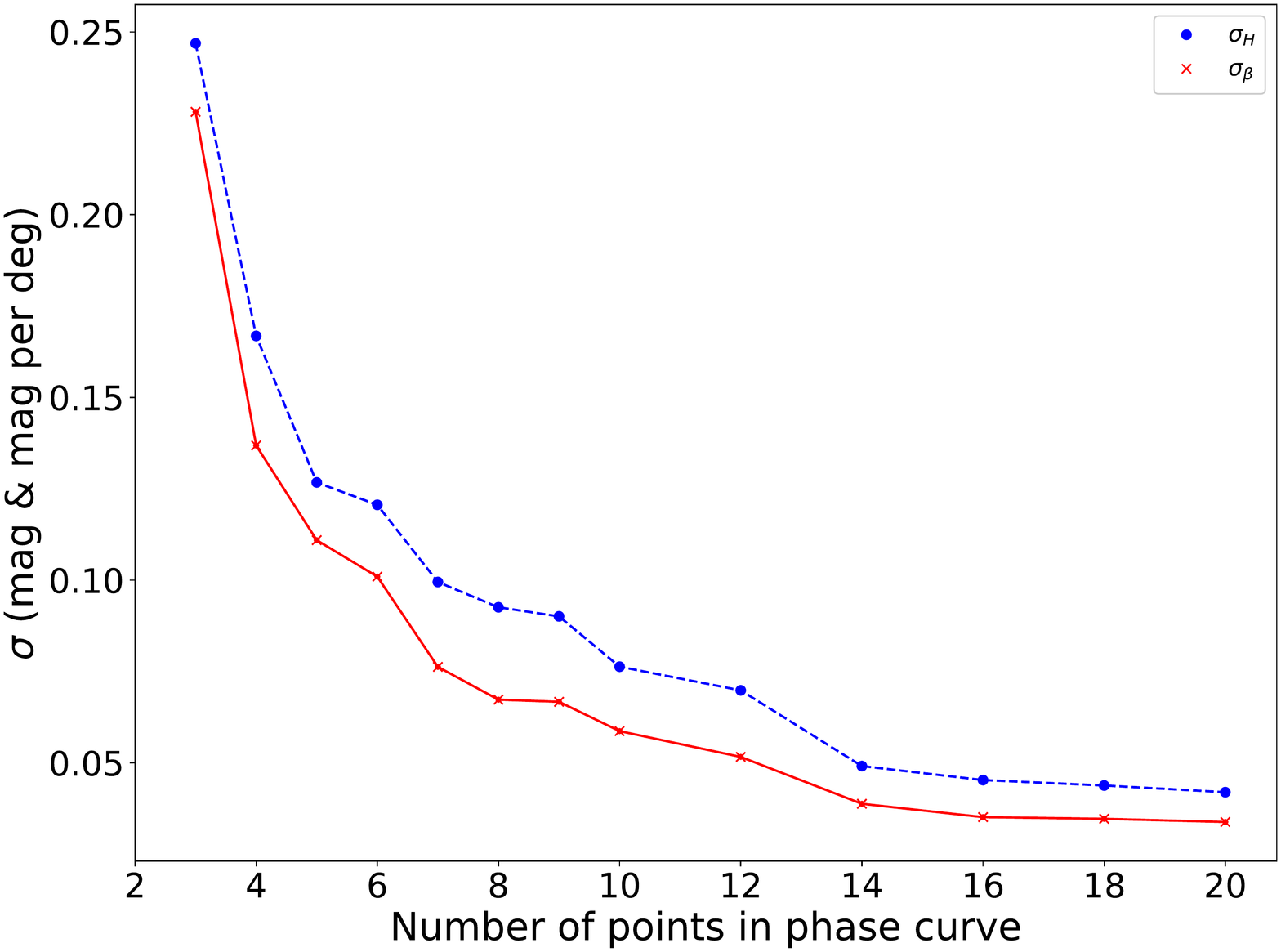}
    \caption{Errors in $H$ (in blue dots) and $\beta$ (in red xs) from simulated data.}
    \label{fig:presicion}
\end{figure}
Nevertheless, it is also good to remember that other systematics, not considered in this simulation, such as zero points offsets or faint sources close to the TNO, could affect the value of $M(1,1,\alpha)$ in ways difficult to predict.
}

\section{Full amplitude vs. half amplitude}\label{appendixa}
To create the phase curves from the collected data we used the data as reported in the literature. In those cases where more than one point per night was reported, for example whenever a light-curve was reported, we took the average value of the night. In other cases, when only one magnitude were reported, we understood it a snapshot at a particular (and unknown) rotational phase.

\begin{figure}
	\includegraphics[width=6cm,angle=90]{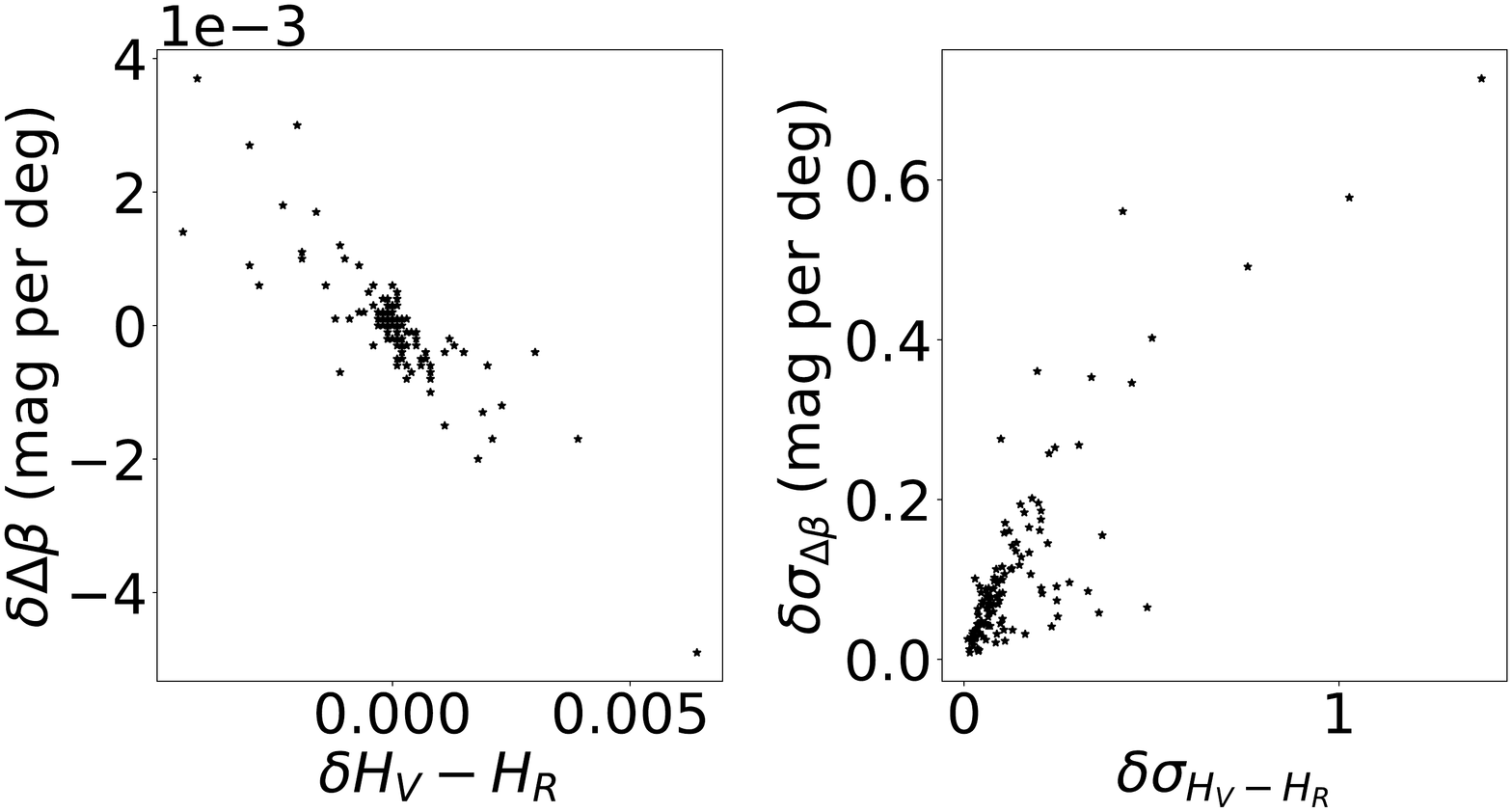}
    \caption{Differences obtained considering $\Delta m$ or $\Delta m/2$ in our processing. Left: Difference between the nominal values of $H_V-H_R$ and $\Delta\beta$, all within 0.6 \%. Right: Difference between the errors, in the larger cases we have errors in colours with 1 magnitude of difference. }
    \label{fig:appendix}
\end{figure}

To deal with the possible effect of the rotational phase we adopted the processing described in Sect. \ref{processing}. But in paper1 and paper2 we assumed, when creating the 100,000 different solutions, that the random numbers could be taken from anywhere between $[-\Delta m,\Delta m]$. While writing this work, we discovered that this overestimated our errors. Therefore, we modified our algorithm to take values from $[-\Delta m/2,\Delta m/2]$, this ensures more precise errors, while leaving the nominal values almost unchanged (Fig. \ref{fig:appendix}).

Using this new approach the median of our errors are $\tilde{\sigma}_{H_V-H_R}=0.09$ and $\tilde{\sigma}_{\Delta\beta}=0.08$ mag per deg.



\bsp	
\label{lastpage}
\end{document}